\documentclass[pra,bibnotes,twocolumn]{revtex4}
\usepackage{graphicx}
\begin{document}
\draft
\def\ds{\displaystyle}
\title{ Pseudo topological insulators}
\author{C. Yuce}
\address{Department of Physics, Anadolu University, Turkey\\ 
Department of Physics, Eskisehir Technical University, Turkey }
\email{cyucem@gmail.com}
\date{\today}
\begin{abstract}
We predict pseudo topological insulators that have been previously overlooked. We determine some conditions under which robust pseudo topological edge states appear and illustrate our idea on the Su-Schrieffer-Heeger (SSH) model with extra chiral symmetry breaking potentials. We discuss that pseudo topological insulating phase transition occurs without band gap closing. 
\end{abstract}
\maketitle

\section{Introduction}

In 1980, the quantum Hall effect in two-dimensional (2D) electron gas at the interface of a semiconductor heterojunction under a strong external magnetic field was observed \cite{klitzing}. In two years, Thouless, Kohmoto, Nightingale and Nijs (TKNN) showed that quantum Hall effect is not only quantum mechanical effect but also a topological effect and hence proposed the concept of topological phases of condensed matter \cite{thoules}. A topologically insulating system is insulating in bulk but has gapless edge states that are immune to local perturbations such as defects and thermal fluctuations. Soon after, Haldane proposed a model to realize topologically nontrivial system without strong magnetic field, which is necessary in the quantum Hall effect to break time-reversal symmetry \cite{halda}. Several years later, Kane and Mele made a breakthrough in the study of topological insulators \cite{kane1,kane2}. They proposed the idea of quantum spin Hall insulator by generalizing the Haldane model. Their model is in fact a graphene model with spin-orbit coupling. As in the quantum Hall effect, symmetry-protected topological edge states appear in the bulk energy gap, but they come in pairs due to time-reversal symmetry and propagate in opposite directions. These edge stares are called helical because of the electron spin-momentum locking. Kane and Mele found a new topological number, so called $Z_2$ number, which is a big finding in the theory of topological insulators. Unfortunately, the spin-orbit coupling in graphene is very weak, and hence this effect cannot be experimentally observed in graphene. Bernevig, Hughes, and Zhang made another breakthrough in 2006 and predicted that CdTe/HgTe/CdTe quantum well can be a 2D topological insulator \cite{bernevig}. This prediction was verified in 2007 \cite{konig}. Since then, the idea of topological phases have attracted great attention and tremendous progress has been made so far.\\
Three discrete symmetries, time-reversal, particle-hole and chiral symmetries, play vital roles in the classifications of topological insulators. Famous "ten-fold way" accounts for topological classification for a given translationally invariant Hamiltonian \cite{symm1,symm2,symm3}. It was realized that a new class of topological insulators unexplored before arise in systems with
 point group symmetries, such as reflection and rotation \cite{andofu}. These are called topological crystalline insulators protected by crystal symmetries. It is well known that $(d-1)$D dimensional gapless boundary states appear in a $d$D topological insulators. Recently, the concept of higher-order topological insulator was introduced \cite{hoti1,hoti2}. Generally speaking, $n$th-order topological insulators $(d-n)$D gapless boundary states. In recent years, non-Hermitian extension of topological insulators and superconductors have attracted great attention especially in photonics community \cite{nonh1,nonh2,nonh3,nonh4,nonh5,nonh6}. In the experiment \cite{nonh2}, stable states that are localized at the interface between two topologically distinct non-Hermitian parity-time symmetric one dimensional photonic lattices with staggered hopping amplitudes were observed through fluorescence microscopy. In the last decade, great progress has been made in our understanding of the theory of topological insulators. We refer the reader for details to some excellent  review papers on this subject \cite{rev01,rev02,rev03,rev04}. \\
In this Letter, we propose a new idea so called pseudo topological insulating phase, which is not compatible with standard classification of topological insulating phase. We show that robust edge states appear in pseudo topological insulators and these states are protected by some symmetries that the system has not. We give some examples to illustrate our idea. Our finding can be easily tested in experiments with current technology.

\section{Formalism}

Consider a general non-interacting first quantized $1D$ Hamiltonian $\ds{\mathcal{H}_0 }$ described by $\ds{(N~{\times}~N)}$ matrix. Assume that this Hamiltonian describes topologically nontrivial lattice with open edges (or at the interface with topologically trivial system). Symmetry protected topological states appear at the edges of the system. Consider an additional potential $\mathcal{V}$ that anticommutes with this Hamiltonian. In fact, this anticommutation condition is exactly the chiral symmetry condition for $\ds{\mathcal{H}_0 }$. We now construct a new Hamiltonian
\begin{equation}\label{olkbf2}
\mathcal{H} =\mathcal{H}_0 +\lambda~  \mathcal{V} ~,~~~~~\{\mathcal{V} ,\mathcal{H}_0\} =0
\end{equation}
where $\lambda$ is a constant. Our aim is to study whether topological edge states of $\ds{\mathcal{H}_0 }$ survive even in the presence of the additional potential. It is clear that $\ds{\mathcal{H} }$ has no longer chiral symmetry and hence one expects that chiral symmetry protected topological edge states disappear. If, for example, the Hamiltonian $\ds{\mathcal{H}_0 }$ is in class $\ds{AIII}$, then the new Hamiltonian $\ds{\mathcal{H} }$ is now in class $\ds{A}$. Note that our system is not restricted to the standart periodic table of topological insulators. Below, we will show that bulk states change with $\mathcal{V}$ while the form of the topological zero energy edge states are preserved even in the presence of $\ds{\mathcal{V} }$. Before going further, we emphasize that there is only one zero energy edge states if $N$ is an odd number while there are two degenerate zero energy edge states if $N$ is even. All other states come in pairs with ${\mp}E$.\\
Consider the eigenvalue equation $\ds{ \mathcal{H}_0 \Psi_{E} =E ~\Psi_{E}}$, where $\ds{ \Psi_{E}}$ are the eigenstates of $\ds{\mathcal{H}_0 }$ with energy $E$. Then the anticommutation condition (\ref{olkbf2}) implies that $\ds{\mathcal{V} \mathcal{H}_0 \Psi_{E} =-\mathcal{H}_0 \mathcal{V} \Psi_{E}=E~ \mathcal{V}\Psi_{E}}$. Assume that $\ds{\Psi_{E}}$ satisfies the equation, $\ds{\mathcal{V}   \Psi_{E} =\nu\Psi_{E}}$, where $\nu=\mp1$, then $\ds{\mathcal{H}_0  \Psi_{E} =-E~\Psi_{E}}$. Therefore, this assumption is true only if $E=0$ and there is no degeneracy at zero energy eigenvalue, which occurs if $N$ is an odd number. (We will study what happens if the zero energy eigenstates are degenerate later.) So, the following equation is satisfied 
\begin{equation}\label{vcbvnrtf2}
\mathcal{V}~  \Psi_{0} =\nu~\Psi_{0}
\end{equation}
where $\nu$ is either $1$ or $-1$ depending on the specific example and $\ds{\Psi_{0}}$ is the zero energy eigenstate of $\ds{\mathcal{H}_0 }$, which is in fact the topological zero energy state. \\
As a result, we can say that $\ds{\Psi_{0}}$ is a simultaneous eigenstate of both $\ds{\mathcal{H}_0 }$ and $\ds{\mathcal{H} =\mathcal{H}_0+\lambda\mathcal{V} }$ with eigenvalues $0$ and $\nu\lambda$, respectively. In other words, the form of nondegenerate zero energy edge state has nothing to do with the extra potential $\ds{\mathcal{V} }$. All other eigenstates of both Hamiltonians are not simultaneous eigenstates. This is interesting since $\ds{\Psi_{0}}$ is well known topological zero energy edge state, while all other eigenstates of $\ds{\mathcal{H}_0 }$ are topologically trivial bulk states. Now two questions arise. i-) Is $\ds{\Psi_{0}}$ an eigenstate of $\ds{\mathcal{V} }$? ii-) Is the simultaneous eigenstate $\ds{\Psi_{0}}$, which is topological for $\ds{\mathcal{H}_0 }$, topological for $\ds{\mathcal{H} }$, too? Below, we answer these questions.\\
Let us start with the first question. The zero energy eigenstate of $\ds{\mathcal{H}_0 }$ is not an eigenstate of $\ds{\mathcal{V} }$ but a linear combination of degenerate eigenstates of $\ds{\mathcal{V} }$. Let $\ds{\phi_n }$ be eigenstates of $\ds{\mathcal{V} }$. The eigenstates are composed of two sets of degenerate eigenstates, $\ds{ \{    \phi_n^{\nu=1}   \}   }$ and $\ds{ \{    \phi_n^{\nu=-1}   \}   }$ with eigenvalues $\nu=1$ and $\nu=-1$, respectively. If $N$ is an odd number, then these two set have not equal number of elements. One can construct the state $\ds{  \Psi_{0}}$ as a linear combination of one set of the degenerate eigenstates in such a way that the Equ. (\ref{vcbvnrtf2}) is satisfied.  \\
Let us now answer the second question. The simultaneous topological zero energy edge state is protected by the chiral symmetry of $\ds{\mathcal{H}_0 }$ and the chiral symmetry is broken in $\ds{\mathcal{H} }$. It is well known that the zero energy eigenvalue of $\ds{\mathcal{H}_0 }$ resists against chiral symmetry protecting disorder. Does the eigenvalue $\nu\lambda$ for the Hamiltonian $\ds{\mathcal{H} }$ resist against the same kind of disorder? It is commonly believed that topological phase disappears with the addition of such a symmetry breaking extra potential term $\ds{\mathcal{V} }$  and hence one expects that the simultaneous eigenstate $\ds{\Psi_0}$ is only topological for $\ds{\mathcal{H}_0 }$. Fortunately, the answer of the second question is Yes despite the common belief! This is striking since the symmetry is broken and the topological number is not compatible for $\ds{\mathcal{H} }$. We call this phase pseudo topological insulating phase. Below we will explain why {\it{nonzero}} edge state of $\ds{\mathcal{H} }$ is robust against disorder that respects the symmetry of $\ds{\mathcal{H}_0 }$. \\ 
Consider a symmetry protecting deformation of the chiral symmetric Hamiltonian: $\ds{\mathcal{H}_0\rightarrow \mathcal{H}_0^{\prime}=\mathcal{H}_0+\epsilon~ V_{SPT}}$, where $\epsilon$ is a small constant and $ {\epsilon }~ V_{SPT}$ is symmetry protecting perturbation potential which preserves the band gap of the original system. The corresponding eigenvalue equation reads $\ds{\mathcal{H}_0^{\prime} \Psi_{E}^{\prime}=E^{\prime} \Psi_{E}^{\prime}   }$. Of special importance is the zero energy state. It is well known that the topological zero energy edge state of $\ds{\mathcal{H}_0 }$ is deformed with $V_{SPT}$, i.e., $\ds{ \Psi_{0}  \rightarrow   \Psi_{0}^{\prime} }$, but its energy remains zero because of the topological robustness ($\ds{\mathcal{H}_0^{\prime} \Psi_{0}^{\prime}=0  }$). Consider now the same deformation for the full Hamiltonian (\ref{olkbf2}): $\ds{\mathcal{H}\rightarrow \mathcal{H}^{\prime}=\mathcal{H}_0^{\prime}+\mathcal{V}=\mathcal{H}_0+\mathcal{V}+\epsilon~ V_{SPT}}$. In this case, the deformation is no longer symmetry protected. Since $ V_{SPT}$ and $\ds{\mathcal{H}_0 }$ has the same symmetry, the anticommutative condition $\ds{ \{\mathcal{V},\mathcal{H}_0^{\prime}\}=0}$ is automatically satisfied. Based on the discussion above the Equ. (\ref{vcbvnrtf2}), we conclude that $\ds{ \Psi_{0}^{\prime} }$ is simultaneous eigenstate of both $\ds{\mathcal{H}^{\prime} }$ and $\ds{\mathcal{H}_0^{\prime} }$. This is interesting since the zero energy eigenstate $\ds{ \Psi_{0} }$ is deformed to $\ds{ \Psi_{0}^{\prime} }$ in the presence of disorder and the extra potential $\ds{\mathcal{V} }$ in (\ref{olkbf2}) has nothing to do with this deformation of zero energy eigenstate. This is good but not sufficient to say that $\ds{ \Psi_{0}^{\prime} }$ is a topological state for $\ds{\mathcal{H}^{\prime} }$, too. We need to show that the energy eigenvalue of this state resists to the potential $ V_{SPT}$. The anticommutative condition $\ds{ \{\mathcal{V},\mathcal{H}_0^{\prime}\}=0}$ implies an equation like (\ref{olkbf2}): $\ds{\mathcal{V}   \Psi_{0}^{\prime} =\nu^{\prime}\Psi_{0}^{\prime}}$, where $ \nu^{\prime}$ is a constant. If $ \nu= \nu^{\prime}$, then we prove the robustness of the energy eigenvalue against $ V_{SPT}$. In fact, it is easy to see that $ \nu= \nu^{\prime}$. This is because there exists degenerate eigenstates of $\ds{\mathcal{V} }$ with eigenvalue $\ds{\nu }$ not $\nu^{\prime}$ as we discussed above. In the presence of $V_{SPT}$, a new linear combination of the degenerate eigenstates $\phi_n$ leads to $\ds{  \Psi_{0}^{\prime}  }$. As a result, the energy eigenvalue $\ds{\nu\lambda}$ remains the same even in the presence of $V_{SPT}$. Note that this topological nonzero energy edge state is not protected by the symmetry of $\ds{\mathcal{H}^{\prime} }$ but that of $\ds{\mathcal{H}_0 }$. So this state is pseudo topological edge state. Note that bulk states change considerably with $\ds{\mathcal{V} }$. \\
So far we have considered the case where there is only one zero energy eigenstate of $\ds{\mathcal{H}_0 }$. Suppose now that there are two degenerate zero energy edge states of $\ds{\mathcal{H}_0 }$, which occurs if $N$ is an even number. Let $\ds{\Psi_{0,A}}$ and $\ds{\Psi_{0,B}}$ be two degenerate zero energy eigenstates of $\ds{\mathcal{H}_0 }$. These two eigenstates are satisfied by $\ds{ \mathcal{V}   \Psi_{0,A} =\nu\Psi_{0,B}  }$ and $\ds{ \mathcal{V}   \Psi_{0,B} =\nu\Psi_{0,A}  }$. So, these can't be eigenstates of the total Hamiltonian $\ds{\mathcal{H} }$. Consider the symmetric and antisymmetric linear combinations $\ds{ \Psi_{0,\mp}=\frac{ \Psi_{0,A}\mp \Psi_{0,B} }{\sqrt{2}}  }$, which are also eigenstates of $\ds{\mathcal{H}_0 }$ with zero energy eigenvalue. Fortunately, $\ds{ \Psi_{0\mp}  }$ satisfies the following equation
\begin{equation}\label{vcrygctf2}
\mathcal{V}  ~ \Psi_{0,\mp} ={\mp}\nu~\Psi_{0,\mp}
\end{equation}
This implies that $\ds{   \Psi_{0,\mp} }$ are eigenstates of the total Hamiltonian: $\ds{ \mathcal{H}  \Psi_{0,\mp}={\mp } \nu \lambda  \Psi_{0,\mp}  }$, which shows us that the degeneracy is lifted by $\ds{\mathcal{V} }$. Note that in the previous case where there is no degeneracy of the zero energy state, the zero energy eigenstate of $\ds{\mathcal{H}_0 }$ and $\ds{\mathcal{H} }$ are the same. But this is not the case when $N$ is even. Let us now discuss what happens if $V_{SPT}$ is introduced into the system. The eigenstates $\ds{ \Psi_{0\mp}  }$ change their forms into $\ds{ \Psi_{0\mp}^{\prime}  }$ but their energy eigenvalues remain the same in the presence of $V_{SPT}$. The new states satisfy the equation $\ds{\mathcal{V}   \Psi_{0\mp}^{\prime} =\mp \nu\Psi_{0\mp}^{\prime}  }$ since these states are eigenstates of $\ds{\mathcal{H}_0^{\prime} }$ with zero energy eigenvalues and the extra potential $\ds{\mathcal{V} }$ anticommutes with $V_{SPT}$. Therefore, we say that the nonzero energy eigenvalues remain the same in the presence of the disorder. \\
Let us now illustrate our idea. We claim that our method works for both Hermitian and non-Hermitian systems since $\lambda$ in Equ. (\ref{olkbf2}) can be a complex number. Consider the well known 1D Su-Schrieffer-Heeger (SSH) Hamiltonian $\ds{\mathcal{H}_0 }$ \cite{ssh} with $\ds{\mathcal{V} =\sigma_z}$ and $\lambda=m+i\gamma$. Therefore $\ds{\mathcal{H}= \mathcal{H}_0 +(m+i\gamma) \sigma_z}$, where $m$ is the mass term and $\gamma$ describes gain and loss in the system when $\gamma>0$ and $\gamma<0$, respectively. Both $m$ and $\gamma$ are real valued constants.  As a special case $m=\gamma=0$, it is well known that finite SSH lattice with open edges can have robust edge states. In a periodical lattice, the total Hamiltonian is given by
\begin{equation}\label{mwedfaz6}
\mathcal{H}=\left(\nu+\omega\cos(k) \right)\sigma_x+\omega\sin(k)\sigma_y+(m+i\gamma)\sigma_z
\end{equation}
where $\vec{\sigma}$ are Pauli matrices, the crystal momentum $k$ runs over the first Brillouin zone, $-\pi<k<\pi$ and the real-valued positive parameters $\ds{\nu>0}$, $\ds{\omega>0}$ are hopping amplitudes. It is easy to see that $\ds{\{\mathcal{V},\mathcal{H}_0\}=0}$. The SSH Hamiltonian has Chiral symmetry but $\ds{\mathcal{H} }$ has no such symmetry. If $m=\gamma= 0$, the winding number is equal to $1$ for $\nu>\omega$ and $0$ for $\nu<\omega$. Therefore, the system is topologically nontrivial for the former case and trivial for the latter. A finite SSH chain with open edges has topological zero energy edge states protected by chiral symmetry when $\nu>\omega$. If $m\neq0$ and $\gamma\neq0$, the winding number is no longer quantized to be an integer and the system is topologically trivial. This can also be seen from the corresponding energy eigenvalues for (\ref{mwedfaz6}). They are given by $\ds{E_{\mp}=\mp\sqrt{ \nu^2+\omega^2+2~\nu~\omega  \cos(k) +(m+i\gamma)^2 }}$, which is symmetrically located around the zero energy. We first study the Hermitian lattice $\gamma=0$ and then we make a straightforward generalization to the non-Hermitian lattice. In the Hermitian case, the band gap never closes and reopens as $\ds{\omega}$ and $\nu$ are tuned for fixed $m$, which is a signature of topologically trivial phase. Although the Hamiltonian (\ref{mwedfaz6}) has no topological phase, it has pseudo topological phase. To check our prediction, we perform numerical calculations for finite lattices with open edges. The parameters we use are $\nu=1.5$, $\omega=0.5$ and $m=1$. We consider $N=20$ and $N=21$ separately, where $N$ is the total number of lattice site. We numerically see that the edge state is not deformed by the extra mass term $\ds{m}$ if $N=21$, while bulk states change considerably with the mass term. The new energy of the edge state is equal to $m=1$. If $N=20$, we see that the degeneracy of the edge states is lifted by the extra mass term and the symmetry between the left and right edge states are broken (they have now different forms and different energy eigenvalues). The eigenvalues of the right and left edge states are exactly equal to $\mp m=\mp1$. These are in total agreement with our predictions. Let us now add disorder and check robustness of these pseudo topological edge states. It is well known that topological edge states in the SSH chain are robust against coupling constant disorder. In standard topological insulator theory, this robustness should be lost if the extra mass term is introduced into the Hamiltonian. In this paper, we predict that they should still be robust because of the existence pseudo topological phase. In our numerical computation, we introduce randomized hopping amplitudes in the lattice. In this way, the coupling constants between neighboring sites become completely independent. The new hopping amplitudes become $\omega\rightarrow\omega+\delta \omega_n$ and $\nu\rightarrow\nu+\delta \nu_n$, where $\delta \nu_n$ and $\delta \omega_n$ are real-valued random set of constants in the interval $[-0.5,0.5]$. We study topological robustness against the disorder for many different random numbers. We find that the energy eigenvalues for the pseudo topological edge states remain to be the same while they change remarkably with the disorder for the bulk states. \\
To see topological phases for various values of tunneling parameter, we parametrize the tunneling parameters as $\ds{\omega=1+0.5 \cos(\Phi)}$ and $\ds{\nu=1-0.5 \cos(\Phi)}$, where modulation phase $\ds{\Phi}$ is another degree of freedom. The Fig.1 plots the spectrum as the parameter $\ds{\Phi}$ is varied. As can be seen, two states with energy $\ds{\mp 1}$ appear in the system (a) unless $\pi/2<\Phi<3\pi/2$ when $N=20$ and a single states with energy $+1$ appears in the whole region (c) when $N=21$. These are pseudo topological states localized around the edges and decay exponentially away from the edges. We add random tunneling disorder into the system, where the random numbers are assigned to each tunneling in the interval $[-0.5,0.5]$. The Fig.1 (b) and (d) plot energy spectra in the presence of the disorder when $N=20$ and $N=21$, respectively. As can be seen, the energy eigenvalues of all bulk states change but the energy eigenvalues of pseudo topological states resist to the disorder. Actually, they resist to the disorder until they merge into the band. \\
\begin{figure}[t]\label{2678ik1}
\includegraphics[width=4.25cm]{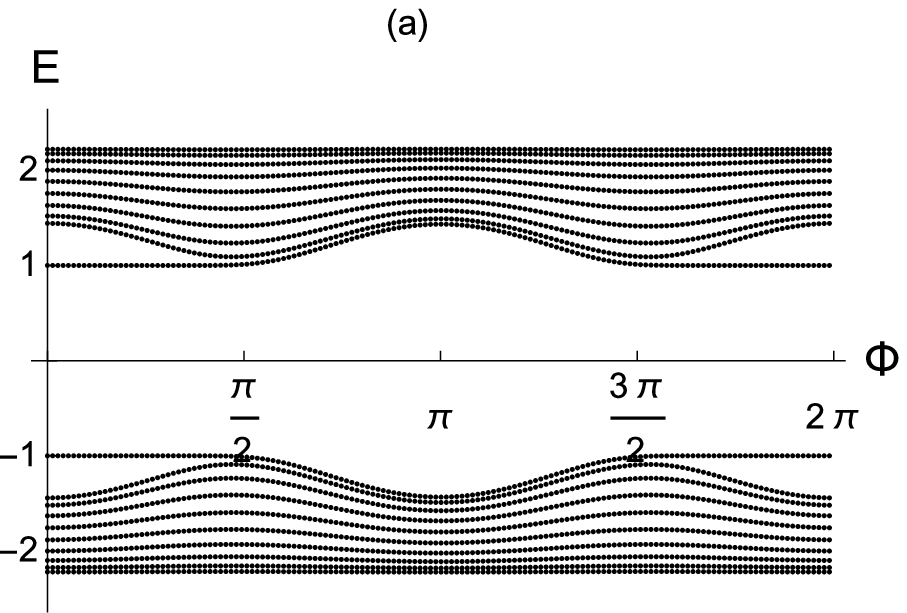}
\includegraphics[width=4.25cm]{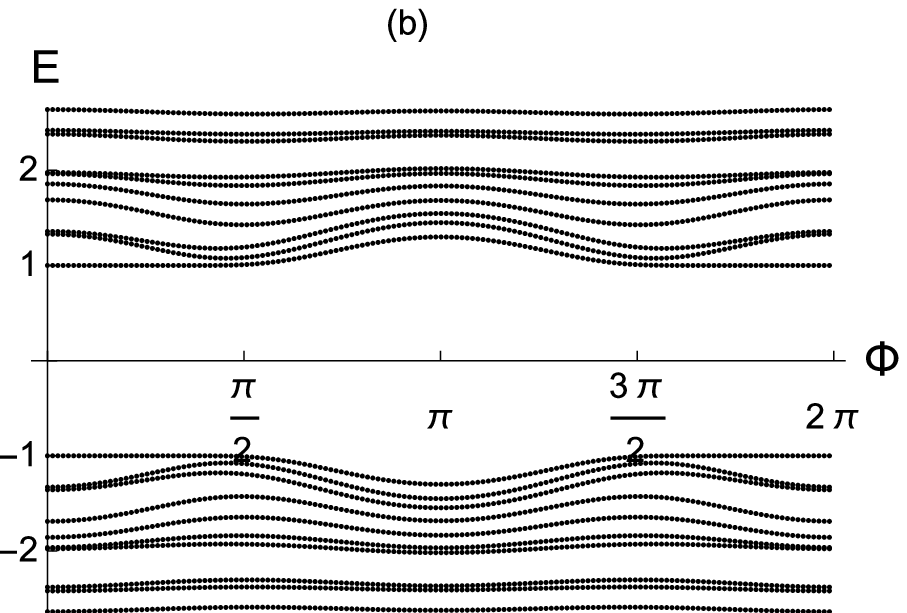}
\includegraphics[width=4.25cm]{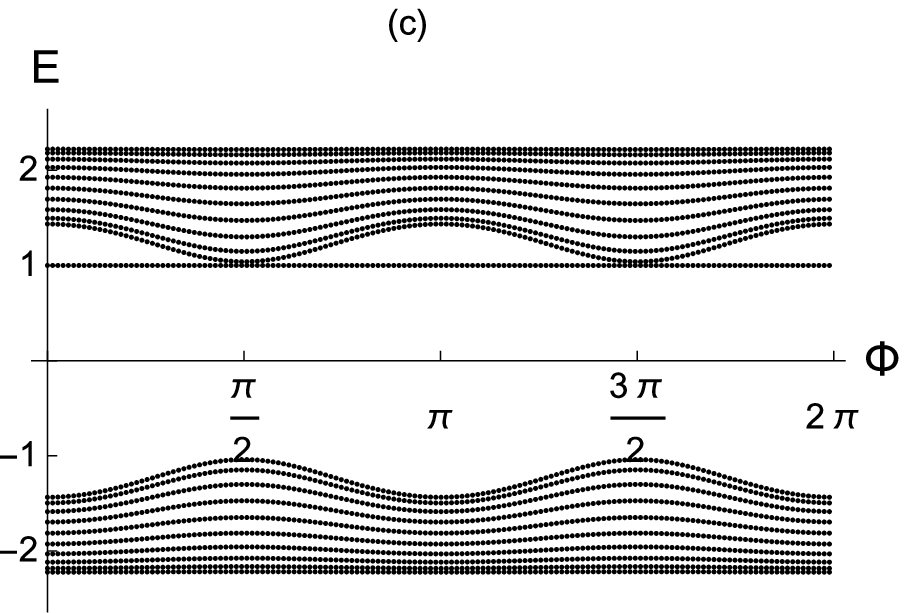}
\includegraphics[width=4.25cm]{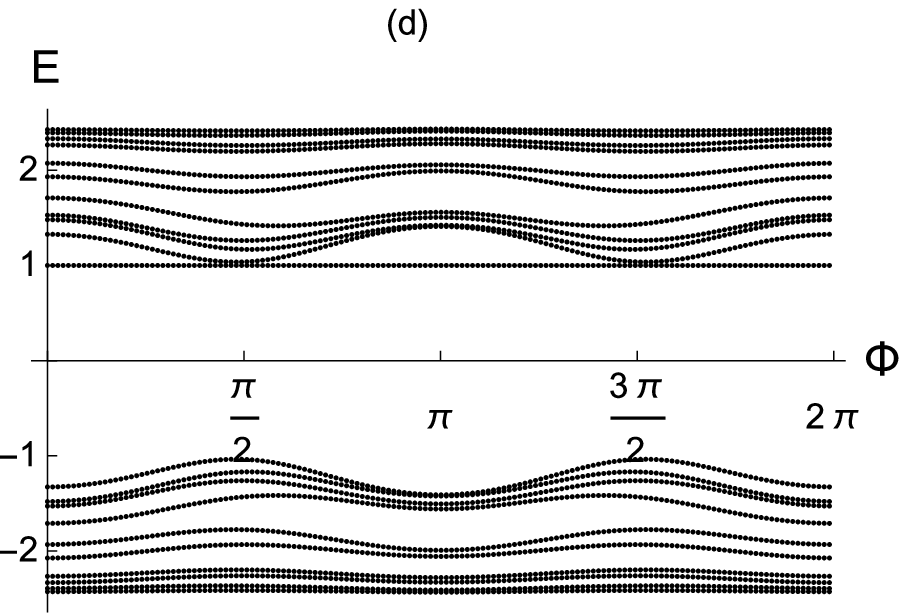}
\caption{ Band structure for the finite chain described by the Hamiltonian (\ref{mwedfaz6}) when $N=20$ (a) and $N=21$ (c). The parameters are $m=1$, $\nu=1+0.5 \cos(\Phi)$ and $\omega=1-0.5 \cos(\Phi)$. Random tunneling disorder is introduced into the system and the Fig. (b) and (d) plot energy spectra in the presence of the disorder when $N=20$ and $N=21$, respectively. In the presence of the disorder, energy eigenvalues of the bulk states change. However energy eigenvalues of the pseudo topological edge states remain the same. }
\end{figure}
Let us now discuss the non-Hermitian extension. The generalization is straightforward. Suppose $\gamma=1$. The energy eigenvalues of the pseudo topological edge states become $\mp (1+i)$ when $N=20$, $(1+i)$ when $N=21$. We check that both the real and imaginary parts of them resist the tunneling disorder. We see that the pseudo topological edge eigenstates exist in the non-Hermitian system, too.\\
Another question arises. Can we find other systems that has pseudo topological phase? We now find another condition for the existence pseudo topological phase. Consider the Hamiltonian $\ds{\mathcal{H}_0 }$ already defined in the Equ. (\ref{olkbf2}) and the chiral operator $\ds{\mathcal{C}=\mathcal{V} }$. We construct a new Hamiltonian $\ds{\mathcal{H} }$ satisfying
\begin{equation}\label{udtvsmfhf2} 
\mathcal{H} =\mathcal{H}_0 +\lambda~\mathcal{U} ~,~~~\{\mathcal{C} ,\mathcal{H}_0\}=0~,~~~[\mathcal{U} ,\mathcal{C}]=0
\end{equation}
where the chiral operator $\ds{\mathcal{C} }$ plays a key role here although it is not explicitly included in the total Hamiltonian. We have discussed above that zero energy topological edge states $\ds{ \Psi_{0}  }$ and $\ds{ \Psi_{0\mp}  }$ satisfy the equations (\ref{vcbvnrtf2},\ref{vcrygctf2}), respectively. Since the two operators $\ds{\mathcal{C} =\mathcal{V} }$ and $\ds{\mathcal{U} }$ commute, these two operators have simultaneous eigenfunctions, too. As a result, the eigenstates $\ds{ \Psi_{0}  }$ and $\ds{ \Psi_{0\mp}  }$ satisfy linear eigenvalue equations for $\ds{\mathcal{U} }$, too. Therefore, $\ds{ \Psi_{0}  }$ and $\ds{ \Psi_{0\mp}  }$ are simultaneous eigenfunctions of both $\ds{\mathcal{H}_0 }$ and $\ds{\mathcal{H} }$. This means that robust pseudo topological edge states exist for the Hamiltonian $\ds{\mathcal{H} }$.\\ 
As an example, consider again chiral symmetric SSH lattice $\ds{\mathcal{H}_0 }$ with $N=21$. Suppose there exists an extra potential whose matrix form is given by $\ds{(U)_{mn}=U_m \frac{1+(-1)^m}{2} \delta_{m,n}}$, where $U_m$ are random numbers. In this way, the potential becomes disordered potential and the total Hamiltonian $\ds{\mathcal{H} }$ is not translationally invariant. According to the standard table of topological insulators, no topological phase occurs. Fortunately, the potential satisfies the conditions (\ref{udtvsmfhf2}) and $\ds{\mathcal{U}   \Psi_{0}=0}$, where $\Psi_0$ is the zero energy eigenstate of the standard SSH lattice with $N=21$. In other words, the zero energy topological edge state of $\ds{\mathcal{H}_0 }$ is not perturbed by this disordered potential although all bulk states changes considerably with it. It is well known that the zero energy topological state of $\ds{\mathcal{H}_0 }$ is robust against chiral symmetry protected deformations. We now study the robustness of the total Hamiltonian $\ds{\mathcal{H} }$ against such kind of deformations. We add random disorder in the hopping amplitudes of the SSH chain and numerically see that zero energy eigenstate of $\ds{\mathcal{H}_0 }$ is robust against such disorder. This shows that pseudo topological edge states appear in the system.\\
Another question arise. Can our approach be generalized to systems that has non-topological zero energy states. One such example is the well-known graphene, which is a two dimensional  honeycomb lattice of carbon atoms. Dirac cones appear in its band structure and hence conducting electrons in graphene move as if they are massless relativistic fermionic particles. Graphene is not a topological insulator since it is a two dimensional gapless system and its Chern number is zero. However, it has protected edge states in the sense that each of the Dirac cones has a Berry phase of $\pi$ and  $-\pi$, respectively. But these edge states are not strictly topological since the total Berry phase vanishes. The edge states in graphene are robust against weak symmetry protected perturbations. Edge states were theoretically predicted and experimentally realized along zigzag edges in graphene \cite{predic1,predic2,grphnedge00,grphnedge01,grphnedge03}. We can construct two new Hamiltonians according to (\ref{olkbf2},\ref{udtvsmfhf2}) where $\ds{\mathcal{H}_0 }$ is the Hamiltonian of the graphene. Therefore, we can say that zero energy eigenstate along the zig-zig edge still survives. In our case, these are pseudo edge states with zero energy.\\
So far, we have discussed pseudo topological phase. One van view the problem in another way. Consider the Hamiltonian $\ds{\mathcal{H}_0 }$. One can study robustness of the topological zero energy edge states of  $\ds{\mathcal{H}_0 }$ against symmetry breaking potentials $\ds{\mathcal{V}}$ and $\ds{\mathcal{U} }$. Normally, we expect that they are robust against symmetry protecting potential. Here, robustness against such symmetry breaking potential $\ds{\mathcal{V}}$ and $\ds{\mathcal{U} }$ occurs. \\
To sum up, we have proposed a new idea so called pseudo topological insulating phase, which arises even if the system has no topological phase. We have given two conditions to observe pseudo topological insulating phase and illustrated our idea on the SSH chain with extra chiral symmetry breaking potentials. We stress that pseudo topological phase transition occurs without band gap closing. Intuitively, we say that one can study other time-reversal and particle-hole symmetric systems to see pseudo topological insulating phase. Such phase may exist in crystalline topological insulators, higher order topological insulators and topological superconductors, too. Therefore, pseudo topological phase are worth studying. Our approach can find some practical applications in photonics and other branches of physics where topological insulating phase can be applied. We have also predicted pseudo topological edge states in time dependent systems.

\end{document}